\documentclass[reprint,prx,amsmath,amssymb,longbibliography,superscriptaddress]{revtex4-1}

\pdfoutput=1

\usepackage{graphicx}
\usepackage{amsmath,latexsym,tabularx}
\usepackage{xcolor}
\usepackage[
  colorlinks=true,
  urlcolor=blue,
  linkcolor=blue,
  citecolor=blue
]{hyperref}
\usepackage{multirow,booktabs}

\newcommand{\eqnRef}[1]{\mbox{Eq. (\ref{#1})}}

\newcommand{\figRef}[1]{\mbox{Fig. \ref{#1}}}

\newcommand{\affilLL}[0]{Lincoln Laboratory, Massachusetts Institute of Technology, Lexington, Massachusetts 02421, USA}
\newcommand{\affilMIT}{Massachusetts Institute of Technology, Cambridge, Massachusetts 02139, USA}
\newcommand{\affilNIST}{National Institute of Standards and Technology, 325 Broadway, Boulder, Colorado 80305, USA}
\newcommand{\unit}[1]{\, \mathrm{#1}}

\begin{document}

\title{Evidence for multiple mechanisms underlying surface electric-field noise in ion traps}

\author{J. A. Sedlacek}
\altaffiliation[Present address:  ]{Honeywell, Golden Valley, MN}
\affiliation{\affilLL}

\author{J. Stuart}
\affiliation{\affilLL}
\affiliation{\affilMIT}

\author{D. H. Slichter}
\affiliation{\affilNIST}

\author{C. D. Bruzewicz}
\affiliation{\affilLL}

\author{R. McConnell}
\affiliation{\affilLL}

\author{J. M. Sage}
\email[]{jsage@ll.mit.edu}
\affiliation{\affilLL}

\author{J. Chiaverini}
\email[]{john.chiaverini@ll.mit.edu}
\affiliation{\affilLL}

\date{\today}

\begin{abstract}

Electric-field noise from ion-trap electrode surfaces can limit the fidelity of multiqubit entangling operations in trapped-ion quantum information processors and can give rise to systematic errors in trapped-ion optical clocks.  The underlying mechanism for this noise is unknown, but it has been shown that the noise amplitude can be reduced by energetic ion bombardment, or ``ion milling,'' of the trap electrode surfaces.  Using a single trapped $^{88}$Sr$^{+}$ ion as a sensor, we investigate the temperature dependence of this noise both before and after \textit{ex situ} ion milling of the trap electrodes.  Making measurements over a trap electrode temperature range of $4$~K to $295$~K in both sputtered niobium and electroplated gold traps, we see a marked change in the temperature scaling of the electric-field noise after ion milling: power-law behavior in untreated surfaces is transformed to Arrhenius behavior after treatment.  The temperature scaling becomes material-dependent after treatment as well, strongly suggesting that different noise mechanisms are at work before and after ion milling.  To constrain potential noise mechanisms, we measure the frequency dependence of the electric-field noise, as well as its dependence on ion-electrode distance, for niobium traps at room temperature both before and after ion milling.  These scalings are unchanged by ion milling.

\end{abstract}

\maketitle

\section{Introduction}

Noise from surfaces is a major source of decoherence for quantum systems, including trapped ions~\cite{theBible,Turchette2000,Brownnutt2015}, superconducting qubits~\cite{Wenner2011,Wang2015}, Rydberg atoms~\cite{PhysRevA.88.043429}, nitrogen-vacancy centers in diamond~\cite{Kim2015,PhysRevB.93.024305}, and nanoelectromechanical devices~\cite{yang2011}.  In trapped ion systems, electric-field noise from surfaces limits the fidelity of quantum logic operations by heating the ions' motion, presenting a challenge for scalable quantum information processing.  It can also introduce systematic shifts in the frequency of trapped-ion atomic clocks~\cite{RevModPhys.87.637,PhysRevLett.118.053002}.  The amplitude of the experimentally measured noise is much larger than would be expected from thermal or technical noise produced by the trap electrodes or external sources.  Because of this unexplained larger amplitude, ion heating from such noise is termed ``anomalous,'' and understanding or mitigating it is of interest both for basic surface science and for applications including quantum information processing. 

The sensitivity of trapped ions to electric-field noise enables their use as exquisitely sensitive surface science probes.  Previous work using trapped ions to sense such noise has shown that treatment of trap-electrode surfaces can reduce the amplitude of the noise at the ion location~\cite{Allcock2011,Hite2012,Daniilidis2014,mcKay2014,McConnel2015}.  These treatments include ion milling, where high-energy atomic ions are directed at the surface in a low-pressure environment; plasma treatment, where a low-energy plasma is created at the surface in a higher-pressure environment consisting of the gases ionized to create the plasma; and laser treatment, where a pulsed laser is directed at the surface.  Factors of reductions in ion heating rates achieved are up to ${\sim}100$ for ion milling~\cite{Hite2012,Daniilidis2014,mcKay2014}, approximately~$4$ for plasma treatment~\cite{McConnel2015}, and approximately~$2$ for laser treatment~\cite{Allcock2011}.  These treatments may be applied \textit{in situ}, i.e. within the same vacuum system as the measurements of noise using individual trapped ions, or \textit{ex situ}, i.e. in a separate system, necessarily requiring a (potentially brief) exposure to ambient atmosphere.  Here we focus on the effect of ion milling, as it has been shown to have the most dramatic effects in reducing electric-field noise in trapped-ion experiments.  Furthermore, we explore the use of \textit{ex situ} ion milling (ESIM) for trap-electrode treatment in particular, as it has the practical advantage that it can be used to treat technologically relevant surface-electrode ion traps without modifications to existing ultra-high-vacuum (UHV) and/or cryogenic systems.

To probe the mechanisms behind anomalous heating, we vary the temperature of the electrode surface.  Prior work has shown a large reduction in anomalous ion heating upon cooling nominally untreated trap surfaces to cryogenic temperatures~\cite{needleTrap,Labaziewicz2008,Chiaverini2014}.  Beyond this reduction, measurements of the exact form of the temperature dependence can also help place limits on potential models~\cite{Labaziewicz2008_2,Bruzewicz2015}.  For instance, models based on the fluctuation of adatoms, either in position or dipole moment~\cite{PhysRevA.87.023421} (or both simultaneously in a correlated manner~\cite{PhysRevA.95.033407}), predict thermally activated noise amplitude, with Arrhenius-type exponential scaling~\cite{Brownnutt2015}. In contrast, models based on thermal fluctuations of charge carriers or atom-polarization in metals~\cite{Brownnutt2015} or insulators~\cite{kumph_NJP_2016} comprising the surfaces predict power-law scalings.  However, to date the temperature dependence of anomalous heating above treated surfaces has not been studied.

In this work, we present measurements of megahertz-frequency electric-field noise above ion-trap electrode surfaces both before and after ESIM as a function of temperature, for two different electrode materials, using a single trapped atomic ion as the sensor.  We also present noise measurements as a function of trap frequency and ion-electrode distance after ESIM for niobium traps at room temperature.  We find that the temperature scalings of the noise before and after ESIM are markedly different, suggesting different mechanisms for anomalous ion heating in the two cases.  With the measured frequency and distance scalings, these data appear to rule out known models for anomalous ion heating (after ESIM) in their current forms.

\begin{figure}[t b !]
\includegraphics[width = 1.0 \columnwidth]{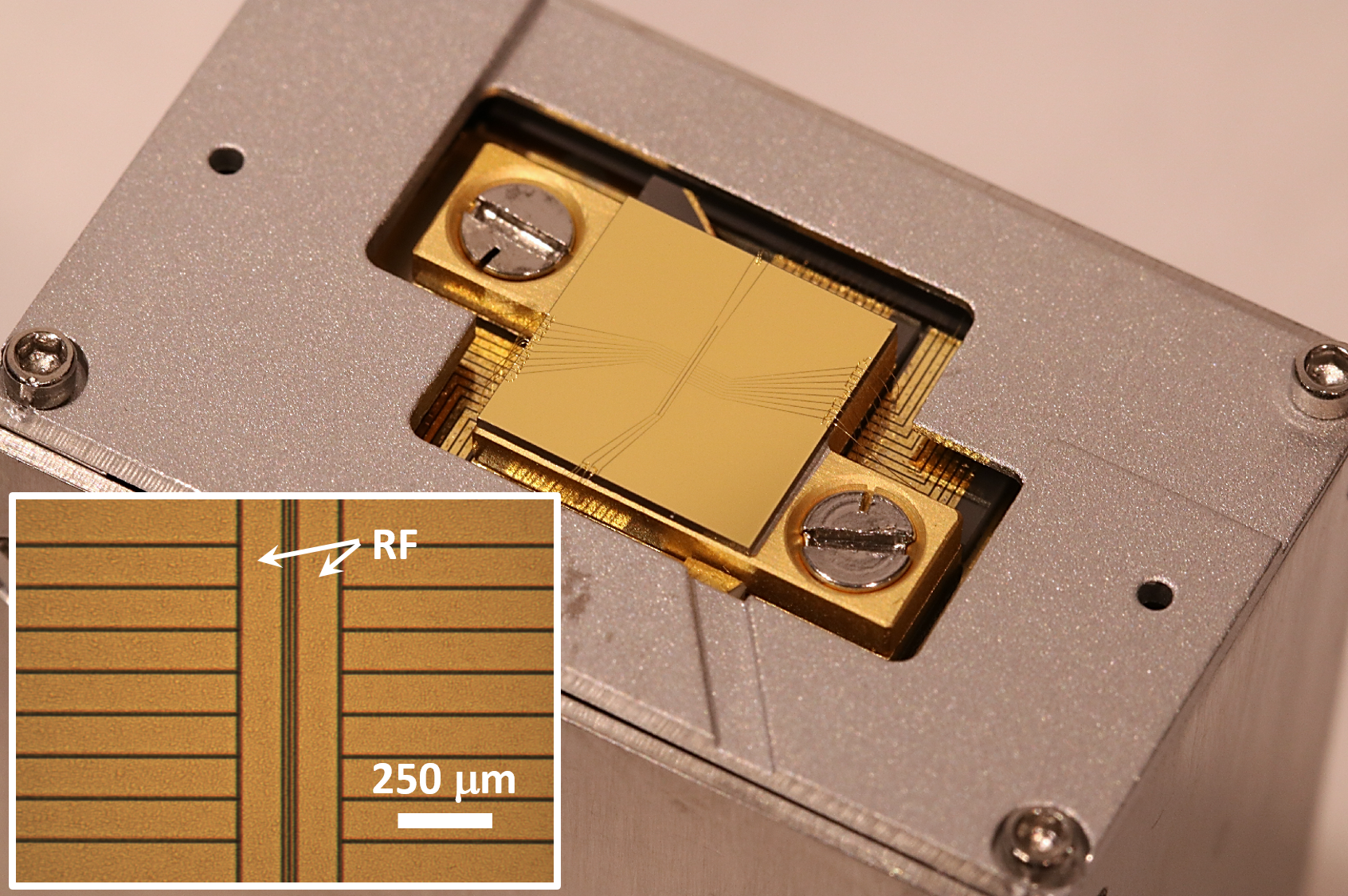}
\caption{Electroplated gold traps used in this work.   The figure is a photograph of the 1-cm-square trap chip attached and wire-bonded to the transfer stage, which is then mounted in either the ion-milling or experimental chamber.  The aluminum cover, below the level of the trap surface, is meant to reduce sputtering of the trap electrode leads and interposer boards underneath.  The inset is a micrograph of the central region of the trap electrodes; the RF-electrode rails are labeled, and all others are DC control electrodes. The ion is trapped $50(1) \unit{\mu m}$ above the center of the linear trap section shown here.  The niobium traps used in this work were of the same design.}
\label{fig:trapPics}
\end{figure}

Electric-field noise near the frequency $f$ of a trapped-ion motional mode with average (thermal) excitation $\bar{n}$ leads to an ion heating rate $\dot{\bar{n}}(\omega,T,d)$ proportional to the electric-field noise spectral density at the ion's location $S_{E}(\omega,T,d)$, where $\omega=2\pi\times f$, $T$ is the electrode temperature, and $d$ is the ion-electrode distance, as

\begin{equation}
\dot{\bar{n}}(\omega,T,d)=\frac{q^{2}}{4 m \hbar \omega} S_{E}(\omega, T, d).
\label{eq:HRvNoise}
\end{equation}

\noindent Here $q$ and $m$ are the ion's charge and mass, respectively, and $\hbar$ is the reduced Planck constant.  Thus, characterization of the ion's motional-state evolution provides a direct measurement of electric-field noise above the surface.  For a single $^{88}$Sr$^{+}$ ion and $f=1.3$~MHz, $S_E\approx [2 \times 10^{-14}$~$(\textrm{V}/\textrm{m})^{2}/\textrm{Hz}\cdot \textrm{s}] \times \dot{\bar{n}}$.  Measuring a heating rate with $1$~quantum/s uncertainty therefore corresponds to electric-field sensing at the $140$~$(\textrm{nV}/\textrm{m})/\sqrt{\textrm{Hz}}$ level.

\section{Experimental System, Surface-Electrode Ion Traps, and Ion-Milling Procedure}

The motional heating measurements are carried out in linear Paul surface-electrode traps using a $^{88}\mathrm{Sr}^{+}$ ion in an apparatus that has been described previously~\cite{Sage2012, Chiaverini2014, Bruzewicz2015}.  Ions are trapped in a UHV cryogenic system which does not require baking of the chamber or trap.  A weak thermal link between the trap chip and the cryostat cold stage allows the temperature of the trap chip to be continuously varied between $4$~and $295 \unit{K}$.  The motional heating rate is measured along the axial direction using sideband spectroscopy~\cite{Chiaverini2014}.  

The linear surface-electrode traps are approximately $7$-$\unit{\mu m}$-thick electroplated (EP) gold, or $2$-$\unit{\mu m}$-thick sputtered niobium, on a sapphire substrate. A micrograph of the electrodes near the center of one of the gold traps is shown in the inset of \figRef{fig:trapPics}, where the layout of the electrodes can be seen.  After fabrication, the traps are coated with photoresist to protect them during dicing and storage.  Traps are rinsed in acetone and isopropyl alcohol, then blown dry with dry nitrogen, prior to wire bonding.  A picture of the trap after wire bonding is shown in the main panel of \figRef{fig:trapPics}.

\begin{figure*}[b t !]
\includegraphics[width = 0.68 \columnwidth]{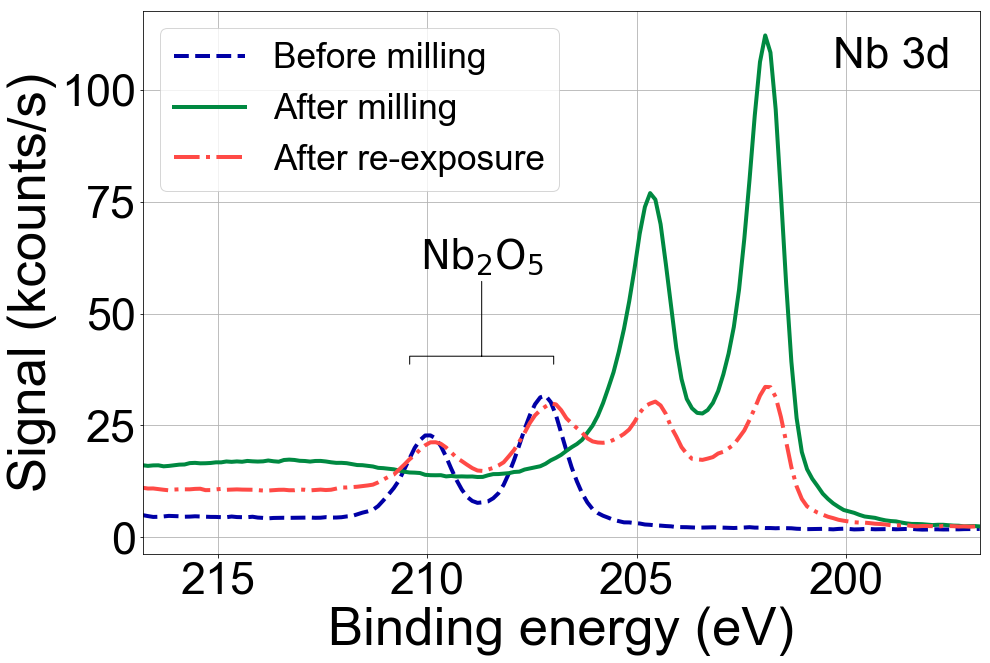}
\includegraphics[width = 0.67 \columnwidth]{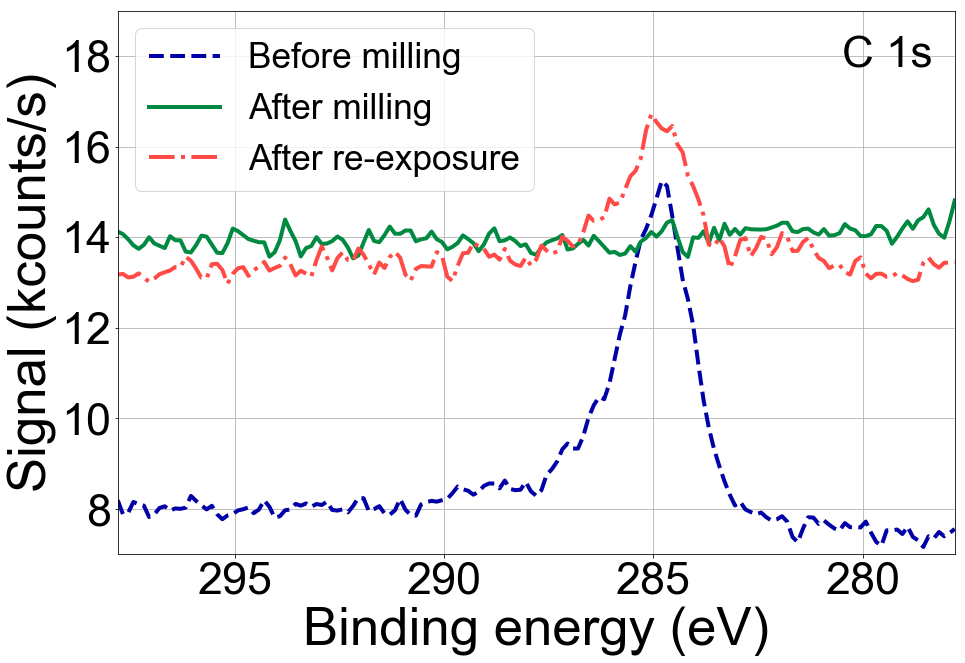}
\includegraphics[width = 0.67 \columnwidth]{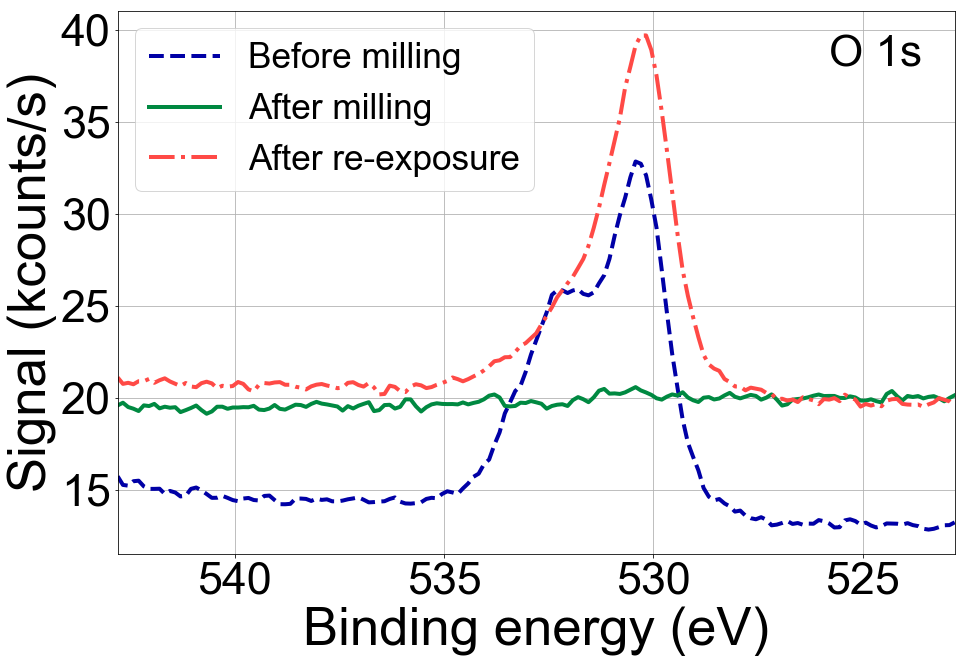}\\
\includegraphics[width = 0.68 \columnwidth]{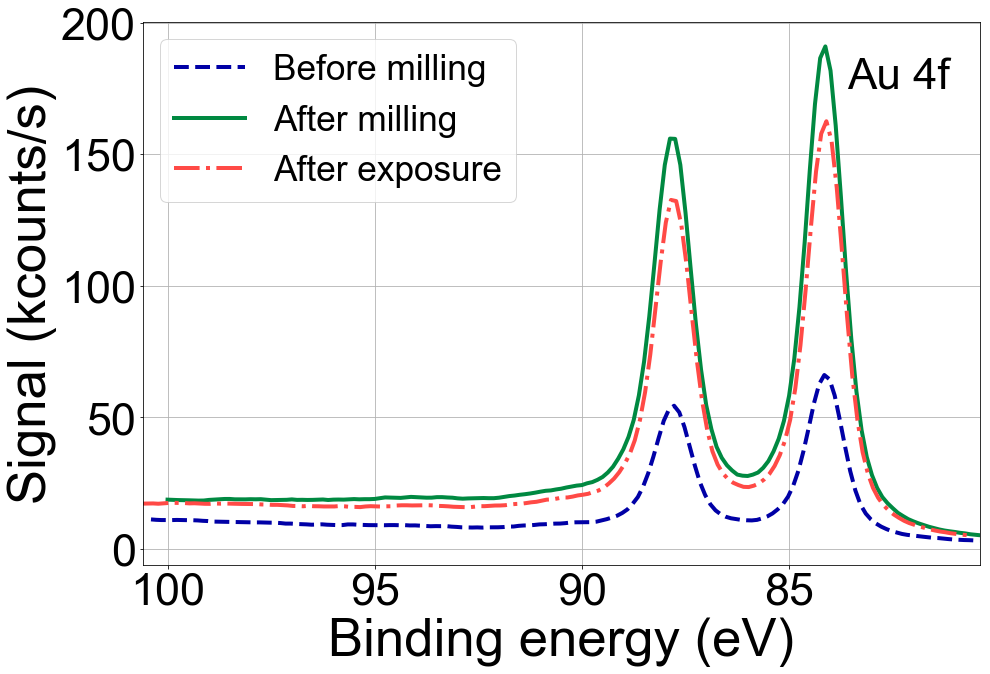}
\includegraphics[width = 0.67 \columnwidth]{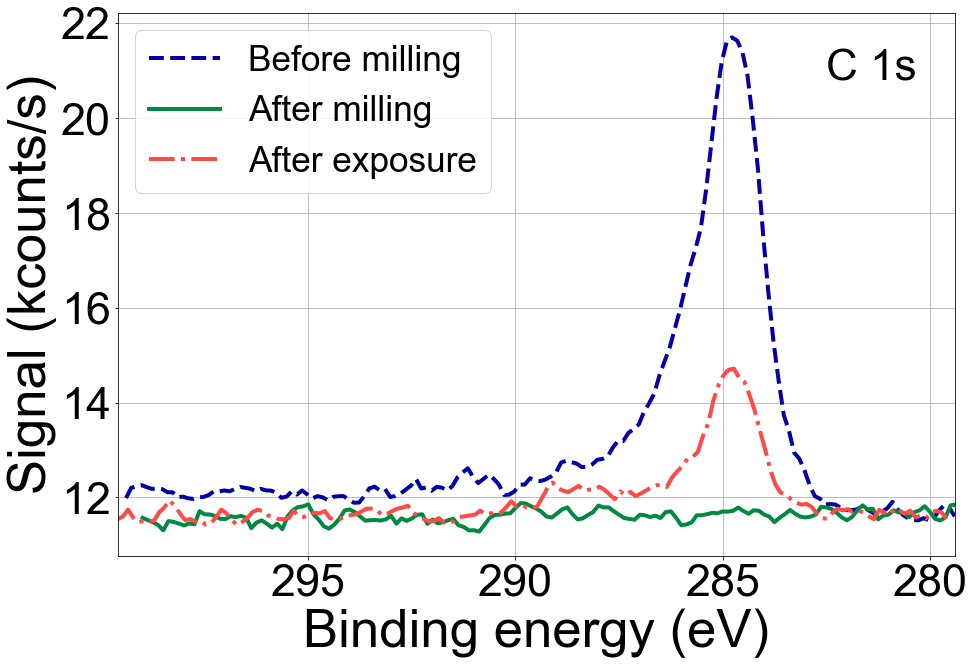}
\includegraphics[width = 0.67 \columnwidth]{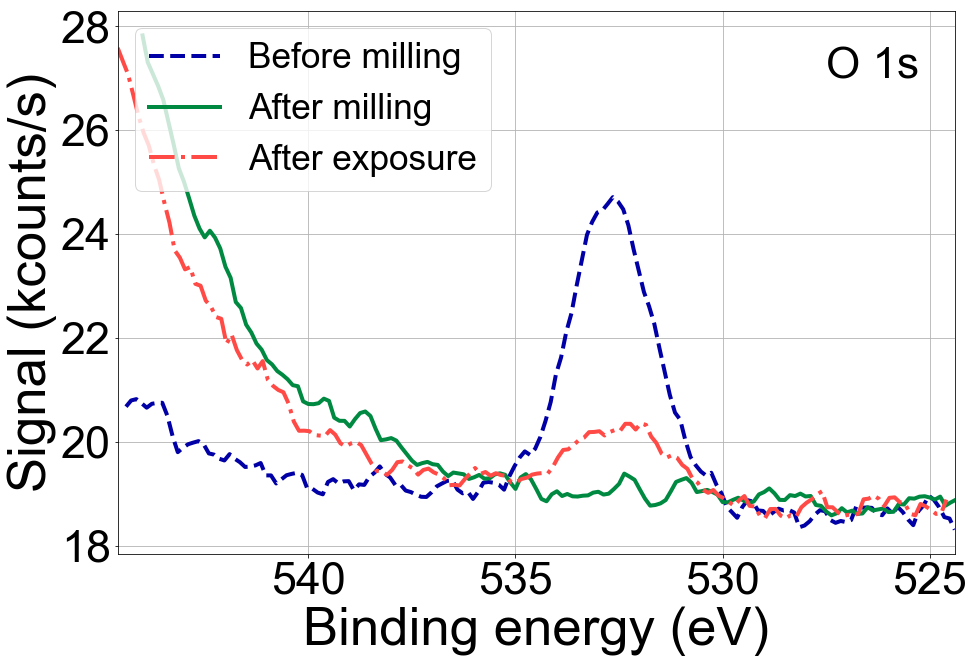}
\caption{X-ray photoelectron spectroscopy (XPS) of the traps used in this work.  Each graph shows data for typical solvent-cleaned chips before milling, after milling in the XPS chamber (2 keV Ar$^{+}$ ions, with a flux density of $6.4\times10^{-1}\unit{(C/m^{2})/s}$, for 2~min), and after a 30~min re-exposure to atmosphere.   Upper (lower) row is Nb (Au).  Left:  Nb~3d (Au~4f) peaks; center:  C~1s peak; right:  O~1s peak.  After milling of Nb, the primarily niobium-pentoxide surface is removed, revealing metallic niobium peaks, while the carbon and oxygen present on the surface, both in carbon-containing compounds and in the oxide, are also eliminated.  After re-exposure, a mixture of niobium and niobium pentoxide is present, and carbon-containing compounds reappear, but at a lower level (visible in both the center and right panels of the top row).  The O peak is the combination of a narrower, lower-binding-energy metallic oxide peak, and a broader higher-binding-energy peak that we associate with hydrocarbons or carbonates.  After milling of Au, the carbon and oxygen present on the surface are eliminated.  We associate these with carbon-containing compounds in part due to the O~peak shift (cf. the O~peak in niobium, upper right panel).  After re-exposure, some contamination returns, but the Au peaks remain single-component in nature, i.e. in contrast to Nb, we see no evidence of oxidation.  The  peak at slightly higher binding energy than the O~peak is the Au~4p peak.  Binding energy is referenced to the adventitious carbon C~1s peak at 284.8~eV.  While the parameters of the milling done in the XPS chamber, particularly the higher Ar$^{+}$ ion flux, lead to a higher material removal rate than the ESIM, we believe these spectra are representative of what would be observed after ESIM since the ion energy and dose are essentially equivalent.  The higher background levels visible in the right two panels in the upper row are primarily due to photoelectrons from Nb atoms which have lost various amounts of energy due to inelastic scattering on their way out of the sample.  There are more such electrons in the ``After milling'' and ``After re-exposure'' cases since there is a higher density of Nb atoms near the surface, due to the smaller amount of oxide and carbon-containing compounds, in these cases.  The high-kinetic-energy edge of these broad backgrounds, appearing just to the left of the Nb peaks, can be seen in the upper left panel at high binding energy.}
\label{fig:xps}
\end{figure*}

Previous ion heating measurements in untreated traps made from Au and Nb have shown similar temperature dependence~\cite{Chiaverini2014}.  When measured at room temperature, traps made from EP gold and treated with ion milling have shown drastic reductions in heating rate compared to before treatment~\cite{Hite2012,mcKay2014}.  While gold does not readily oxidize with exposure to atmosphere, and furthermore has been shown to not gain oxygen after ESIM and air exposure~\cite{mcKay2014}, niobium forms a few-nanometer-thick oxide when exposed to air.  X-ray photoelectron spectroscopy (XPS) on a niobium trap chip (performed in a separate, dedicated apparatus~\footnote{XPS parameters were as follows for the measurements presented here.  The x-rays are from a monochromated Al-K$\alpha$ source with energy 1486.6~eV; the x-ray beam diameter is 200~$\mu$m.  Charge compensation was performed for all scans, and the analyzer pass energy was approximately 60~eV.}) shows that milling produces a pure metallic surface that acquires a partial coverage of niobium pentoxide after a 30~min exposure to air, with a mixture of pairs of metallic and oxide peaks visible in the spectroscopy (See~\figRef{fig:xps}, top row). In contrast, similar XPS measurements on a gold trap chip show only pure metallic components in the region of the gold peaks before milling, after milling, and after re-exposure (See~\figRef{fig:xps}, bottom row).  Though carbon and oxygen are present before milling and after re-exposure in this case, this observation is consistent with carbonaceous contaminants and not with a metallic oxide.  Our exploration of these two materials in this study is motivated by their similar behavior prior to ion-milling treatment despite their difference in oxidation susceptibility.

The ESIM is carried out in a separate vacuum chamber. Inside the milling chamber, an ion sputtering gun (OCI Vacuum Microengineering~\footnote{Commercial products are identified in this paper for informational purposes only. Such identification does not imply recommendation or endorsement by the National Institute of Standards and Technology, nor is it intended to imply that the products identified are necessarily the best available for the purpose.}) is mounted perpendicular to the trap surface, so that accelerated Ar$^+$ ions impact the trap chip at normal incidence.  The parameters for the ion milling used in this work are $2 \unit{keV}$ ion beam energy, $5 \times 10^{-6} \unit{Torr}$ background partial pressure of Ar, and an ion flux density of $3 \times 10^{-2} \unit{(C/m^{2})/s}$.  The ion flux density was determined by measuring the ion current through the trap electrodes.  These parameters lead to a material removal rate of approximately 0.64(9)~nm/min as measured via profilometry over a step in the gold film between ion-milled and masked sections.  From the expected 2~keV sputter yield~\cite{matsunami_1984} and measured Ar$^{+}$-ion flux density, we calculate an expected material removal rate of 0.61~nm/min, equal, within error, to the measured value.  From a similar calculation for niobium, we expect the material removal rate to be 0.24~nm/min, but it was not measured independently.  Each trap is treated for a variable amount of time before being exposed briefly to the ambient laboratory air and transferred to the main chamber.  The trap is exposed to atmosphere for ${\sim}1 \unit{h}$ during the transfer.

After the initial cleaning of the traps with acetone and isopropyl alcohol, no additional such cleaning was performed before each trap was subsequently inserted into the ion trap apparatus or the milling chamber.  Initial heating rates of the axial vibration mode at a frequency $f$ of approximately 1.3~MHz were measured using an unmilled trap; the trap was then removed from the main experimental chamber and mounted in the milling chamber. Additional heating rate measurements were performed after transferring the trap back to the experimental chamber.  This process  of milling and heating-rate measurement was repeated, and subsequent milling treatments were performed with the trap being exposed to atmosphere during each transfer.

\section{Temperature Dependence Before and After Ion Milling}

\begin{figure}[b t !]
\includegraphics[width = 0.97 \columnwidth]{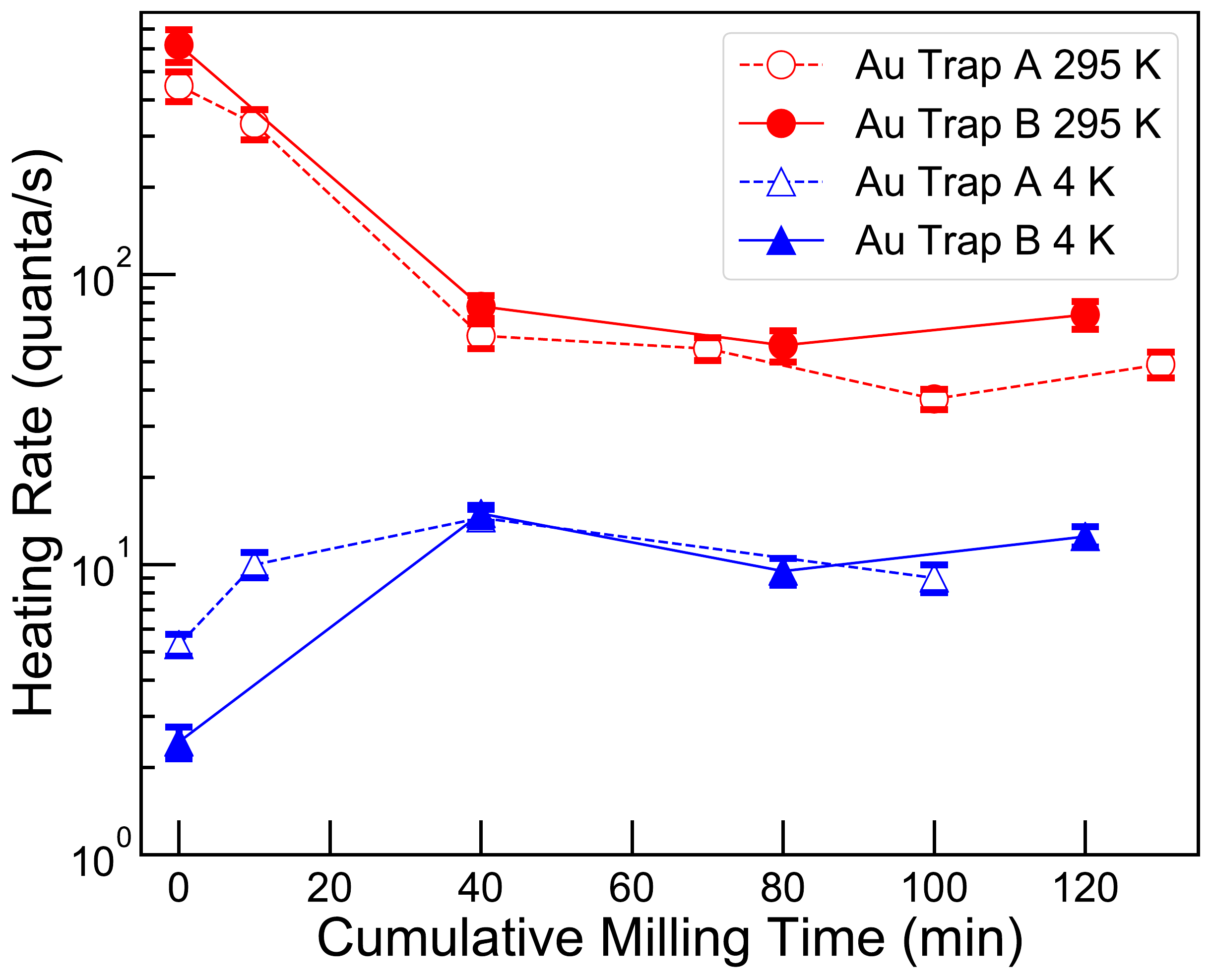}
\caption{The heating-rate plateauing behavior after increasing amounts of \textit{ex situ} ion milling for two nominally identical gold traps (labeled A, depicted by the open symbols, and B, depicted by the closed symbols) with electrodes at 295~K and 4~K; here the heating rate is measured on the axial mode at 1.3~MHz.  For each time step, each trap was exposed to air and transferred to and from the milling chamber.  The milling time represents the total integrated time that the trap was milled.  With the exception of duration, every milling step used nominally the same parameters: $5 \times 10^{-6} \unit{Torr}$ $\mathrm{Ar}$, an ion beam energy of $2 \unit{keV}$, and an ion flux density of $3 \times 10^{-2} \unit{(C/m^2)/s}$.  The lines connecting data points are intended as a guide to the eye.  Similar data (not shown) was acquired for a niobium trap (trap C), with a plateau time in that case of approximately 80~min.}
\label{fig:increase}
\end{figure}

\begin{figure}[t b !]
\includegraphics[width = 1.01 \columnwidth]{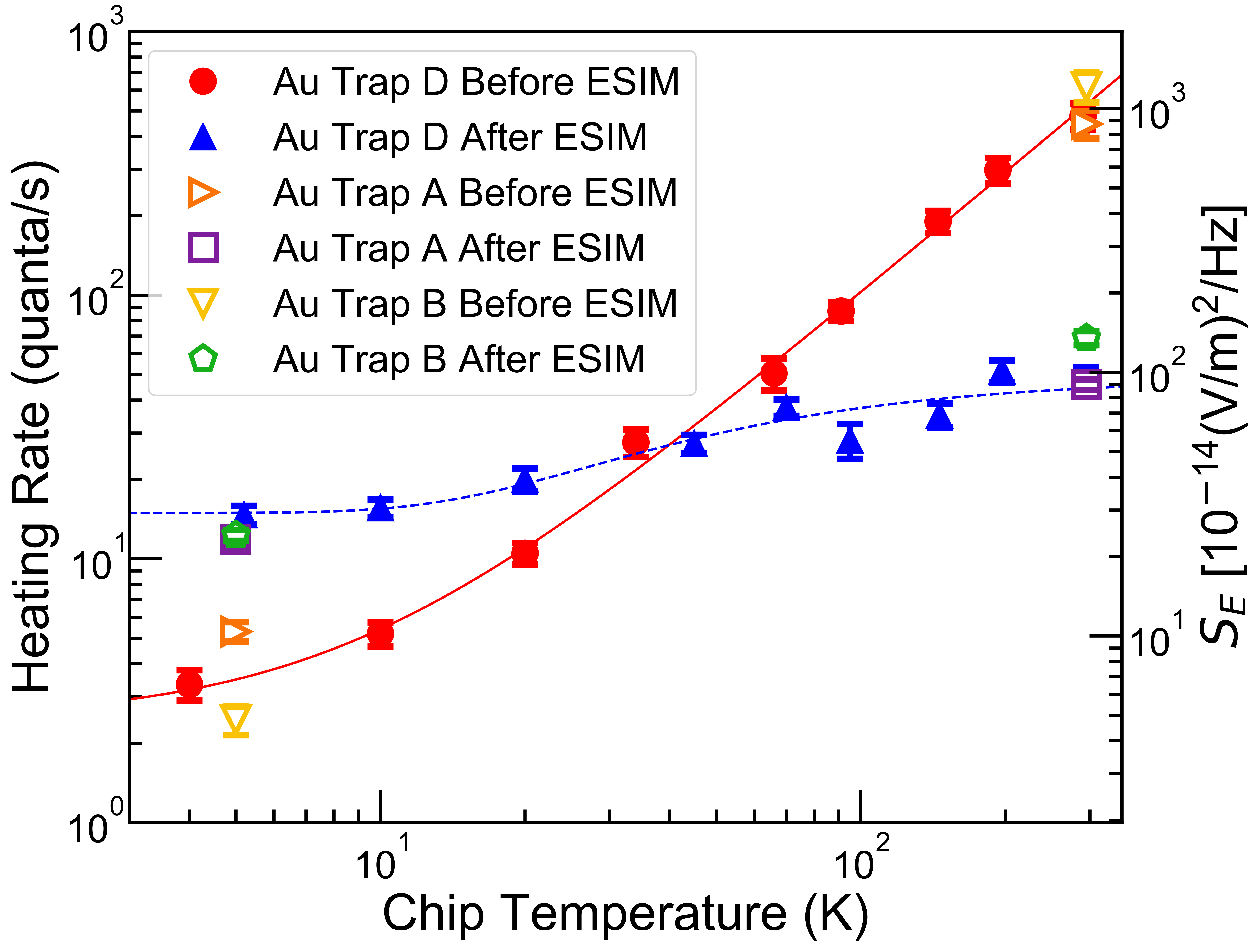}\\
\includegraphics[width = 1.01 \columnwidth]{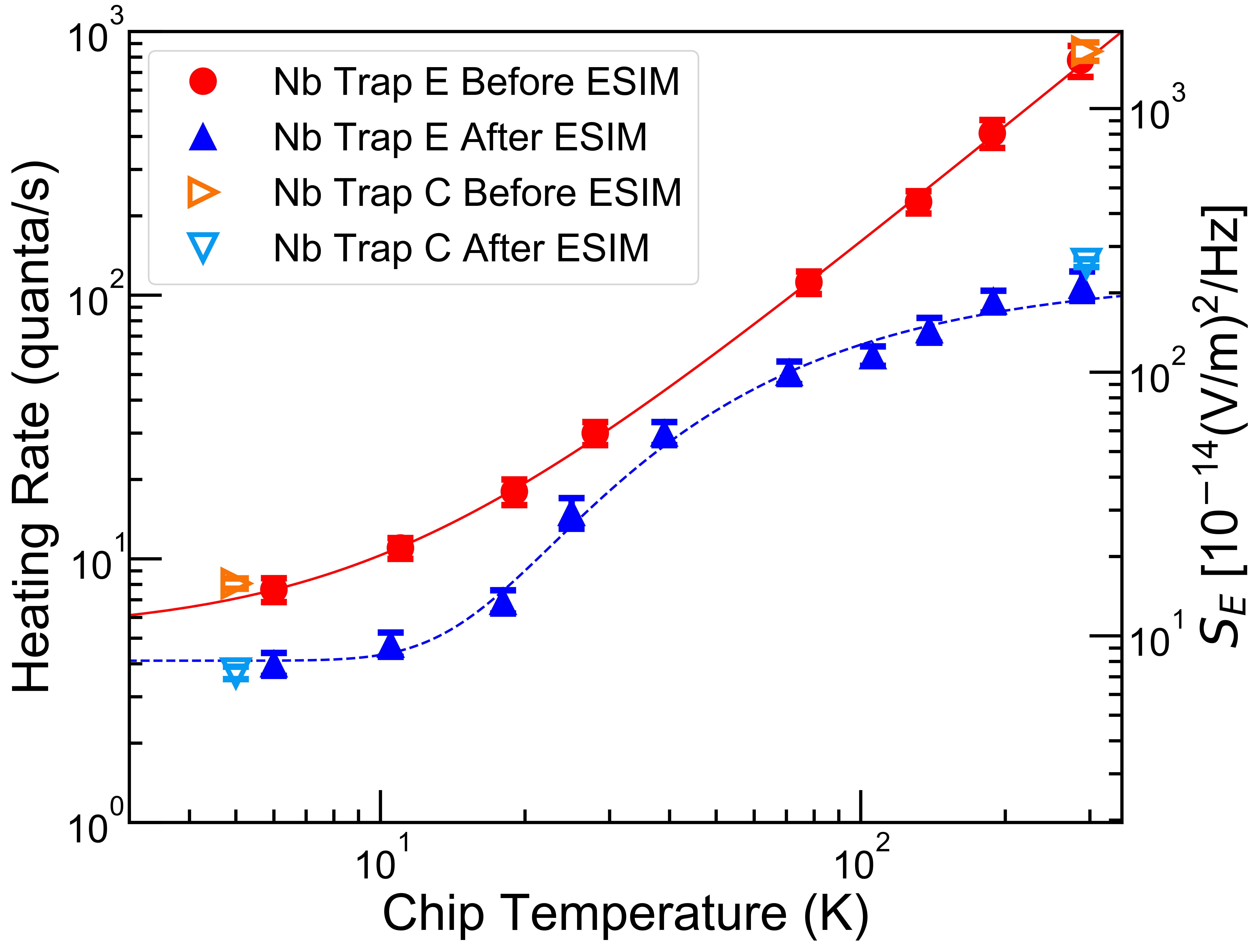}
\caption{Comparison of temperature dependence of heating rates measured on the 1.3~MHz axial mode before and after \textit{ex situ} ion milling for both electroplated gold (top, milling time 60~min) and sputtered niobium (bottom, milling time 100~min).  The solid (red) lines are fits of the round points to \eqnRef{eq:em}, and the dashed (blue) lines are fits of the triangular points to \eqnRef{eq:ah}.  Key fit parameters for the top [bottom] graph:  a pre-ESIM scaling exponent $\beta$ of 1.53(6) [1.48(7)], a pre-ESIM power-law thermal-activation temperature scale $T_{\mathrm P}$ of 9(2)~K [10(2)~K], and a post-ESIM Arrhenius activation temperature scale $T_0$ of 41(9)~K [63(4)~K].  The post-ESIM heating rate in the gold trap was also measured at a trap frequency of 660~kHz (not shown), yielding an Arrhenius fit with a temperature scale $T_0$ of 51(11)~K, equal within error to that determined from the 1.3~MHz heating rate data.  Initial and ESIM plateau data for traps A and B (C) are also displayed in the top (bottom) figure to show trap-to-trap variability for each material.  The right axes are translated to electric-field noise spectral density via \eqnRef{eq:HRvNoise}.}
\label{fig:beforeAndAfterMilling}
\end{figure}

After confirming that one round of ESIM treatment reduced heating rates at $295 \unit{K}$ for both gold and niobium traps, we performed subsequent treatment on the same traps to map out the change in heating rates with further milling.  Concurrently, we measured the effect on the heating rate near $4 \unit{K}$.  The results for two different gold traps (labeled A and B) are shown in \figRef{fig:increase}.  In both cases, the heating rates plateau after ${\sim} 40$~min of milling.  The plateau behavior appears after ${\sim} 80$~min of milling for niobium, not shown (trap~C; see \figRef{fig:beforeAndAfterMilling} for initial and plateau values).  Perhaps surprisingly, the amount of time required to reach the plateau region corresponds to significant material removal:  approximately 25~and 20~nm for gold and niobium, respectively.  Since ESIM roughens the surface while redepositing sputtered material as it proceeds, however, the complete removal of hydrocarbon and oxide layers of 2~to 10~nm in thickness may require substantial additional milling time. For the traps of both materials, the room temperature heating rates at the plateau are lower than the heating rates of untreated traps by a factor of ${\sim} 10$. Interestingly, however, the heating rates near $4 \unit{K}$ are increased in gold traps.  The time to reach a plateau was the same for both temperatures, which indicates that the mechanism responsible for the change was the same in both cases.

To further investigate this change in temperature dependence, additional traps of each material were used to measure heating rates at various temperatures from $4$~to $295 \unit{K}$ before and after ESIM.  The results are shown in \figRef{fig:beforeAndAfterMilling}.  The pre-ESIM heating rates ([red] solid, circular points in the top and bottom panels) are fit to a power law~\cite{Labaziewicz2008_2,Bruzewicz2015},

\begin{equation}
\dot{\bar{n}}(T) = \dot{\bar{n}}_0 \left[1+\left(\frac{T}{T_{\mathrm P}}\right)^\beta \right],
\label{eq:em}
\end{equation}

\noindent where $\dot{\bar{n}}_0$ is the temperature-independent heating rate, $T_{\mathrm P}$ is the thermal activation temperature, $\beta$ is the high-temperature power law exponent, and $T$ is the temperature of the electrodes. After lowering the electrode temperature from ${\sim} 295 \unit{K}$ to ${\sim} 4 \unit{K}$, the heating rate is reduced by a factor of ${\sim} 100$, which is typical in our system for a variety of trap materials and fabrication methods~\cite{Chiaverini2014,Bruzewicz2015}.  The scaling exponents and activation temperatures are the same within error for the gold and niobium traps, also consistent with previous measurements, e.g.~\cite{Bruzewicz2015}, where power-law scaling exponents in the range of~1.5 to~1.6 were measured.

However, after milling, very different behavior is observed (See \figRef{fig:beforeAndAfterMilling}, [blue] solid, triangular points in both panels).  First, the functional form is changed; the heating rates appear to approach an asymptote at both high and low temperatures, with the positive curvatures at high temperature pre-ESIM becoming negative.  Second, the values of the heating rates of gold and niobium now differ significantly; in the case of the gold trap, the heating rates are higher than the initial measurements for trap temperatures below $50 \unit{K}$, whereas for the niobium trap, the post-ESIM heating rate is lower over the whole temperature range.  Moreover, the data after milling do not fit well to \eqnRef{eq:em} with a power law exponent in the range of all previous measurements (i.e. $1.5<\beta<4$)~\cite{Labaziewicz2008_2,Chiaverini2014,Bruzewicz2015}, but rather show an activated behavior characteristic of Arrhenius scaling, 

\begin{equation}
\dot{\bar{n}}(T) = \dot{\bar{n}}_0 + \dot{\bar{n}}_T\, e^{-T_0/T}.
\label{eq:ah}
\end{equation}   

\noindent Here $\dot{\bar{n}}_T$ is the high-temperature contribution to the heating rate and $T_0$ is the Arrhenius activation temperature.

As comparison of $\chi^{2}$ goodness-of-fit values cannot strictly and generally be used to determine which of multiple models best represents a given data set, we use the Bayesian information  criterion (BIC)~\cite{schwarz_1978} for model comparison.  The BIC is a score based on the likelihood function and a penalty for the number of parameters used; the latter component serves to avoid over-fitting and to promote model parsimony.  The BIC is an increasing function of the error variance and of the number of model parameters.  When comparing multiple models, the one with the lowest BIC is preferred, and the more the difference between the preferred model and the others, the more support there is for the lowest-BIC model (the posterior probability of the model given the data is proportional to $e^{-\textrm{BIC}/2}$); the difference in BIC can therefore be assessed absolutely, and any difference is positive evidence for the lowest-BIC model.  Differences in BIC larger than approximately 6 are considered strong evidence, while differences larger than approximately 10 are considered very strong evidence~\cite{kass_and_raftery}.

In comparing the power-law and Arrhenius models (Eqs.~\ref{eq:em} and~\ref{eq:ah}), there is very strong evidence for power-law behavior in the pre-ESIM data for both materials.  On the other hand, the post-ESIM data provides very strong evidence for Arrhenius behavior in niobium and positive evidence for Arrhenius behavior in gold (See Table~\ref{tab:model_comp}~\footnote{BIC Values are calculated using the Mathematica software package, ver. 11 (Wolfram Research, Inc.), via the error variance during nonlinear curve fitting.}).  While the evidence for the Arrhenius behavior over power-law behavior in post-ESIM gold is not strong, we point out that the best-fit power-law exponent $\beta$ for these data is 0.36(14), significantly different from the pre-ESIM value and from all measured previously or expected theoretically~\cite{Brownnutt2015}.

\begin{table}[b !]
\caption{Bayesian information criterion (BIC) values used for model comparison of temperature dependence in pre- and post-milled heating rate measurements (data from traps D and E in \figRef{fig:beforeAndAfterMilling}).  The model with the lower BIC value is preferred (indicated in bold-face type for each condition).  The BIC difference value $\Delta$BIC, the score of the lower-BIC model subtracted from that of the higher-BIC model, gives a measure of evidence for the lower-BIC model (probability $\propto e^{-\textrm{BIC}/2}$).  In this case, there is very strong evidence for power-law behavior prior to ESIM and slight positive to very strong evidence, depending on material, for Arrhenius behavior after ESIM.}
\begin{ruledtabular}
\begin{tabular}{llccc}

\multicolumn{2}{c}{Condition}   & \multicolumn{2}{c}{BIC values} &  \multirow{2}{*}{$\Delta$BIC} \\
\multicolumn{2}{c}{and Material} & Power law      & Arrhenius    &                               \\

\hline
\multirow{2}{*}{Pre-ESIM}  & Au & \textbf{59}  & 134            & 75  \\
                           & Nb & \textbf{55}  & 103            & 48  \\
\multirow{2}{*}{Post-ESIM} & Au & 56.8         & \textbf{55.5}  & 1.3  \\
                           & Nb & 95           & \textbf{55}    & 40  \\ 

\end{tabular}
\end{ruledtabular}
\label{tab:model_comp}
\end{table}

For the gold trap we find the best Arrhenius-model fit for $T_0$ is 41(9)~K (see \figRef{fig:beforeAndAfterMilling}, top), while for the niobium trap, $T_0$ is 63(4)~K (both measured at 1.3~MHz trap frequency).  We have measured the heating rate in the same gold trap at a 660~kHz trap frequency as well, and in that case we also see Arrhenius behavior with $T_0=51(11)$~K. Detailed temperature dependence at other trap frequencies has not yet been measured in niobium traps after ESIM.  However, the data from niobium traps F and G (presented below), which show distance and frequency dependence at room temperature post-ESIM, can be extrapolated to estimate the heating rate at 1.3 MHz and 50 um ion-electrode distance.  The extrapolated values are consistent with the measured room-temperature post-ESIM heating rates for niobium traps C and E (\figRef{fig:beforeAndAfterMilling}b).  Moreover, while detailed temperature dependence was measured on only one trap of each material (traps D and E), data taken pre- and post-ESIM at room temperature and near 4~K using the three other traps (A, B, and C) are all consistent with the altered temperature dependence described above.

These observations are indicative of different mechanisms for anomalous ion heating before and after ESIM, i.e. for solvent-cleaned compared to milled surfaces.  Moreover, the hydrocarbons that adsorb during air exposure after ESIM do not contribute to electric-field noise in the same manner as those present after solvent cleaning; even though the milling is performed \textit{ex situ} in this case, its effect is not nullified by re-adsorption of carbon-containing compounds from the atmosphere.  Similarly, re-adsorption of oxygen and carbon in UHV conditions after ion milling has been previously seen to not increase ion heating rates~\cite{Daniilidis2014}.  We note that Arrhenius behavior has been observed once before in a single trap~\cite{Labaziewicz2008_2}; in the measurements performed here with ESIM, the temperature-dependent behavior change was observed in all traps studied.  Also, the existence of temperature dependence after ESIM in the experiments presented here suggests that they are not limited by technical noise. 

\begin{figure}[t b !]
\includegraphics[width = 0.90 \columnwidth]{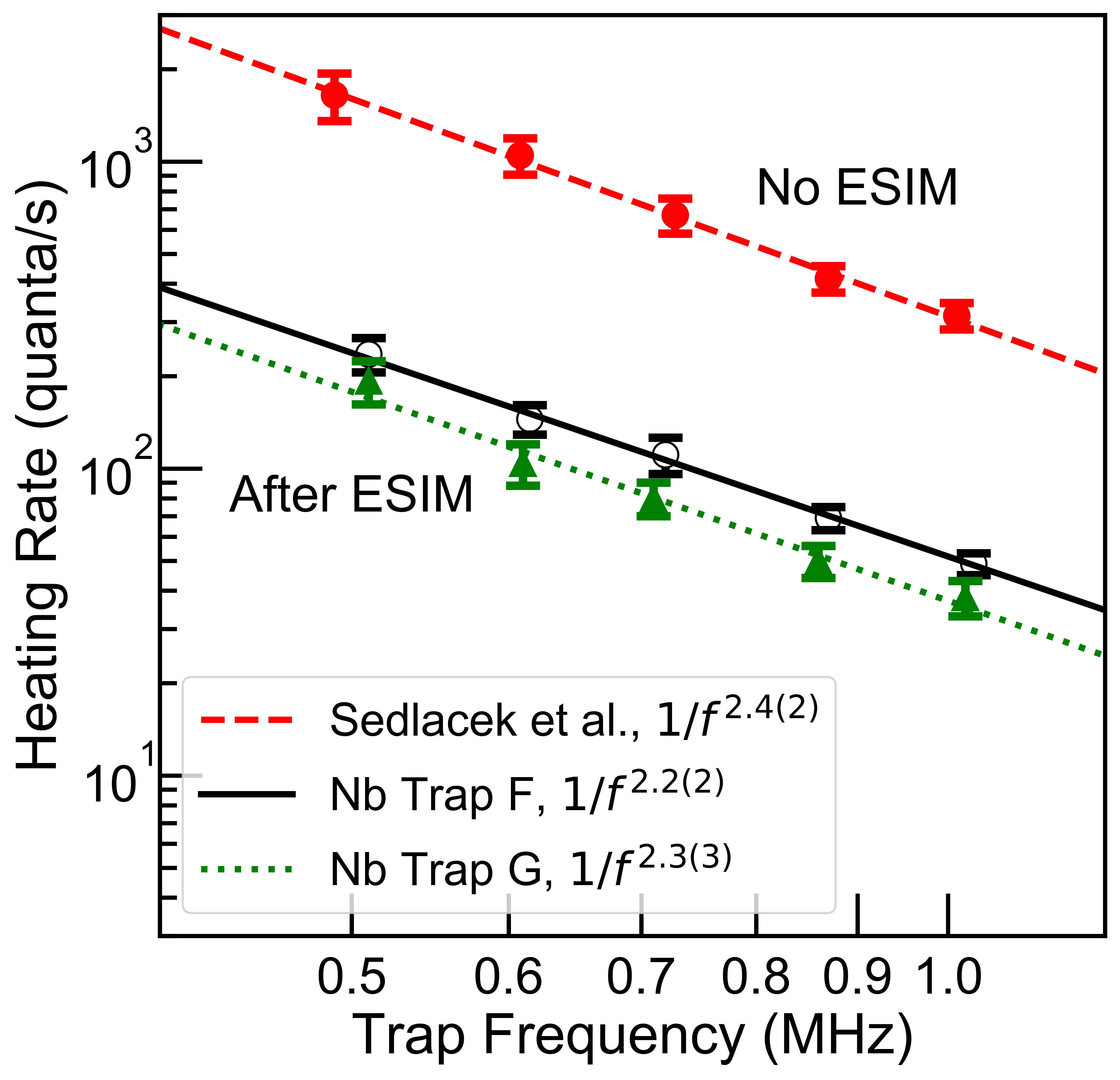}\\
\includegraphics[width = 0.93 \columnwidth]{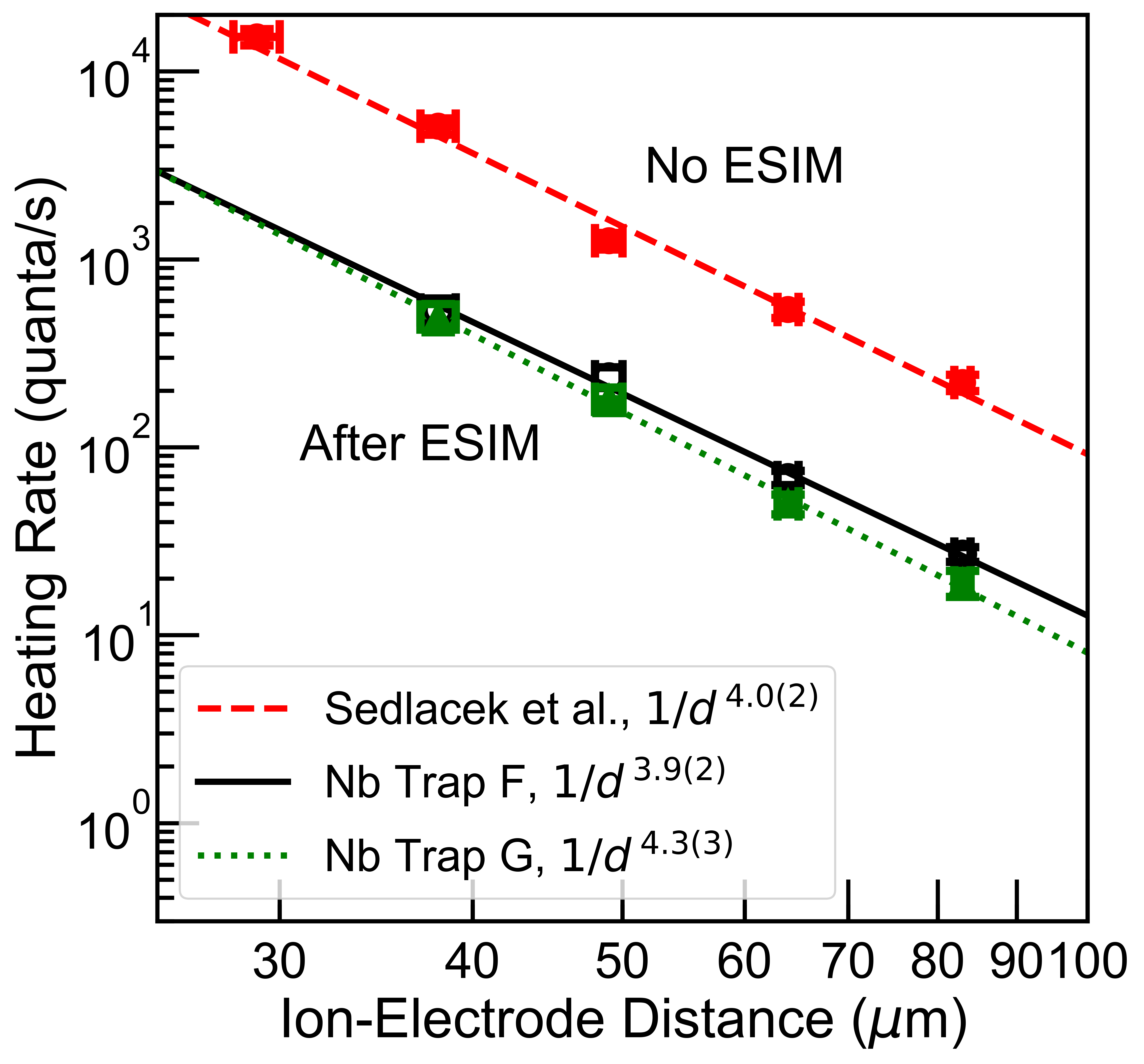}
\caption{Frequency (top panel) and distance (bottom panel) scaling of ion heating rates in Nb traps at room temperature before and after \textit{ex situ} ion milling (ESIM).  The round, solid (red) data points are from a previous measurement~\cite{sedlacek2018} and were taken without ESIM.  Using two traps of the same design (labeled F and G) as in that work, data were taken after ESIM [open circles (black) and triangles (green)].  The ion-electrode distance was $64\unit{\mu m}$ for the measurement as a function of frequency (top), and the trap frequency~$f$ was $860 \unit{kHz}$ for the measurement as a function of distance.  The lines are power-law fits with exponents as depicted in the legend.  The traps used for these measurements are of a different electrode design than those used for the temperature-dependent measurements in this work (see~\cite{sedlacek2018} for details), though they were made in the same process run on the same wafers.}
\label{fig:freqdist}
\end{figure}

Of the leading theoretical models proposed to explain anomalous ion heating, the power law scalings of the temperature dependence for the pre-ESIM measurements follow the lossy dielectric model \cite{kumph_NJP_2016} most closely.  Noise, under this hypothesis, originates from the dissipative nature of any dielectric film covering the electrode metal; electric-field noise from this source is distinct from, but analogous to, the Johnson noise of a metal, though here it is based on thermally driven fluctuations in a polarizable material.  The model predicts a linear scaling $(\beta = 1)$ of the heating rate with~$T$, while we measure $\beta \approx 1.5$ for both materials.  This model also predicts the $1/d^{4}$ distance scaling (for ion-electrode distance $d$) measured in planar surface traps \cite{sedlacek2018,wunderlich2018}, and its $1/f^{2}$ scaling is consistent with widely measured heating-rate frequency scaling~\cite{Brownnutt2015} (c.f. also~\figRef{fig:freqdist}).  We note that an extension of the lossy dielectric model to include temperature dependence of the dielectric constant and loss tangent may alter the temperature dependence to agree more closely with our measured scaling; this is plausible given that the loss tangents of many insulators decrease as temperature decreases.      

Turning now to the post-ESIM measurements, Arrhenius behavior of the temperature scaling is predicted by both the fluctuating dipole (FD) model~\cite{PhysRevA.84.023412,PhysRevA.87.023421} and the adatom diffusion (AD) model~\cite{Brownnutt2015}.  The FD model is based on phonon-induced dipole-moment fluctuation of adatoms, and its predictions include heating rate scalings of $1/f$ with frequency (i.e. a $1/f^{0}$ scaling, or frequency independence, of the electric-field noise power spectral density $S_{E}$) in the range relevant to ion trap frequencies (${\sim}1$~MHz), and of $1/d^{4}$ with ion-electrode distance.  The Arrhenius-type behavior is predicted at temperatures below an effective temperature $T_{\textrm{FD}}$ set by vibrational modes of adatoms bound to surfaces, estimated to be approximately 50~to 100~K.  Above this temperature, the noise is expected to fall as ${\sim}1/T$~\cite{PhysRevA.84.023412}, or to grow as a power law in temperature with an exponent of approximately~$2.5$~\cite{Brownnutt2015}.  The AD model, which is based on field-fluctuations due to the dipole moments of adatoms moving along the surface, predicts Arrhenius temperature scaling over the whole temperature range, with frequency and distance heating-rate scalings of  $1/f^{3}$ and $1/d^{6}$, respectively.  An extension to the AD model (EAD) which considers adatoms diffusing over patches of the surface, where they take on varying dipole moments such that spatial-temporal correlations appear in the noise~\cite{Brownnutt2015}, also predicts  Arrhenius temperature scaling, $1/f^{2.5}$ heating-rate frequency scaling, and $1/d^{6}$ distance scaling for motional modes parallel to the planar-trap surface, as in the case of the axial mode measured here.  See Table~\ref{tab:theories} for a summary of the model predictions and the scalings observed in this work.

We note that the Arrhenius scaling with temperature predicted by these two models differs at low temperature.  While the electric-field noise is expected to be exponentially suppressed for temperatures below $T_{\textrm{FD}}$ under the FD model, the AD and EAD models predict a temperature-independent level of noise at the lowest temperatures, due to diffusion driven by quantum tunneling~\cite{Brownnutt2015}.  The post-ESIM data presented here also shows an approach to a temperature-independent level at low temperature.

\section{Trap-Frequency and Ion-Electrode Distance Scalings}

In light of the altered temperature dependence after ESIM, we measured frequency and distance scaling after ESIM using niobium traps at 295~K, both to determine if these scalings are also affected, and to constrain the FD, AD, and EAD models, as their predicted frequency and distance scalings are different.  While the frequency dependence was not seen to change after milling in previous work with gold~\cite{Hite2012}, niobium has not been explored, and no measurements of distance scaling after ESIM have been reported previously.  Variable-height linear traps~\cite{sedlacek2018} made in the same sputtered-niobium process were used for these post-ESIM measurements; since the multiple zones of this trap design are spread over a region of a few square millimeters around the chip center, these traps were milled for 120~min to ensure every site was milled to the plateau (cf.~\figRef{fig:increase} and surrounding discussion).  Results are shown in \figRef{fig:freqdist} where they are plotted with the measurements from~\cite{sedlacek2018}, performed using a niobium trap chip that had not undergone ESIM.  Unlike the temperature scaling, neither the $1/f^{{\sim} 2.4}$ frequency scaling nor the $1/d^{{\sim} 4}$ distance scaling measured before ESIM is significantly changed after ESIM.

Thus, while the temperature scaling seen here is supportive of the FD, AD, and EAD models, we see discrepancies with each of them when taking all the ESIM data together (See Table~\ref{tab:theories}).  The FD model predicts the observed distance scaling, but does not fit the frequency dependence well---the current theory requires unrealistically heavy or loosely bound adsorbates~\cite{PhysRevA.87.023421} to bring the frequency scaling into the observed range for standard ion trap parameters; in this range, however, the temperature scaling matches well (Arrhenius with a high-$T$ asymptote).  The AD and EAD models both make accurate predictions for the frequency scaling behavior, the EAD slightly more so; the distance scaling, however, is not predicted well.  In the latter case, where patch geometry is relevant, a more detailed incorporation of the adatom-patch dynamics could potentially lead to different distance dependence.  We hope that the material-dependent Arrhenius scaling and additional constraints suggested by these observations will motivate avenues for further understanding of the relevant mechanisms through modification of these, or other, microscopic theories.

\begin{table}[b !]
\caption{Predicted and observed scalings (measured in this work) of ion heating rates for vibrational modes parallel to the surface-electrode trap surface.  Electric-field noise scaling is the same as heating-rate scaling except in the case of frequency, where $1$ should be added to the scaling exponent (cf. \eqnRef{eq:HRvNoise}).  ($\dagger$) The temperature dependence of the noise in the lossy dielectric model may be strengthened by additional temperature dependence of the material loss tangent, typically an increasing function of temperature in this range.  (*) The temperature dependence for the fluctuating dipole model is predicted to be Arrhenius-like up to an effective temperature scale of a few tens of kelvin; above this, the noise is expected to scale either as $1/T$ or $T^{{\sim}2.5}$.}
\begin{ruledtabular}
\begin{tabular}{llll}

\multirow{2}{*}{Model}   & \multicolumn{3}{c}{Predicted $\dot{\bar{n}}$ Scalings}  \\
                         & Temperature      & Freq.    & Distance    \\

\hline
 Lossy dielectric~\cite{kumph_NJP_2016}       & $T$ ($\dagger$) & $f^{-2}$      &  $d^{-4}$    \\
 Fluct. dipole~\cite{PhysRevA.84.023412,PhysRevA.87.023421}      & $e^{-T_0/T}$ (*)& $f^{-1}$      &  $d^{-4}$    \\
 Adatom diffusion~\cite{Brownnutt2015}        & $e^{-T_0/T}$    & $f^{-3}$      &  $d^{-6}$    \\
 Extension to diffusion~\cite{Brownnutt2015}  & $e^{-T_0/T}$    & $f^{-2.5}$    &  $d^{-6}$     \\
\\

 Condition & \multicolumn{3}{c}{Observed $\dot{\bar{n}}$ Scalings}\\
\hline
 Pre-ESIM              & $T^{1.51(4)}$           & $f^{-2.4(2)}$ & $d^{-4.0(2)}$ \\
 Post-ESIM             & $e^{-T_0/T}$            & $f^{-2.2(2)}$ & $d^{-4.0(2)}$ \\
                       & $T_{0}^{\rm Au}=45(7)$~K &               &               \\
                       & $T_{0}^{\rm Nb}=63(4)$~K &               &               \\

\end{tabular}
\end{ruledtabular}

\label{tab:theories}
\end{table}

\section{Discussion}

The temperature scaling results suggest particular methodologies for mitigation of ion heating rates.  In particular, for traps operated at room temperature, ESIM provides approximately a factor of ten reduction in heating rates for gold or niobium; a milling step prior to chamber installation should be performed in this case.  For traps operated at low temperatures, ESIM seems useful for niobium, but counterproductive for gold; in the latter case this step should be avoided.  One caveat to these general comments is that high-temperature system bakes, which may be required to reach UHV after ESIM and chip installation in non-cryogenic systems, were not performed in this work.  Such baking may potentially reduce or alter the effect of ESIM.

Untreated traps have previously been shown to lead to material-independent anomalous heating behavior~\cite{Chiaverini2014}, suggesting that similar contaminants, from processing or solvent cleaning and air exposure, are the dominant sources of electric-field noise across materials.  The emergence of material dependence after ESIM, however, gives hope that the exploration of different materials will lead to more basic understanding of the mechanism behind anomalous heating of treated surfaces, since it provides a new experimental variable.  In particular, the observed increase in electric-field noise in post-ESIM gold surfaces over untreated surfaces seen here at low temperatures suggests that the temperature-independent component of the underlying noise mechanism in treated gold is not only larger in the megahertz regime than that in niobium, but also larger than the material-independent noise mechanism due to solvent or other-hydrocarbon residue seen on as-fabricated samples.  Linking this observation to a unique property of gold could be accomplished by comparison of several surface materials after ESIM.  Potentially more practically useful in the near term, material dependence suggests further reduction of heating rates through electrode material or morphological choice in combination with ESIM.  Both avenues make clear the importance of a re-investigation of electric-field noise as a function of electrode material, with likely impacts beyond trapped ions, touching on many areas where surface-generated noise limits performance.  Moreover, our observation of drastically different behavior of electric-field noise before and after surface ion milling reiterates the utility of individual ions as sensors for furthering our understanding of surface phenomena. 

\textit{Note added---}During the review process, we became aware of related measurements of ion heating as a function of temperature, in this case above room temperature for unmilled electrodes~\cite{noel2018_arxiv}.  We can use a subset of the data analyzed in the present work for comparison to the thermally activated fluctuator (TAF) model.  This model is suggested by the authors of~\cite{noel2018_arxiv} to produce frequency-scaling power-law exponents in agreement with their high trap-temperature electric-field noise measurements.  Our measurements of ion heating rates as a function of temperature in unmilled Nb traps (the data presented in Fig.~\ref{fig:beforeAndAfterMilling}) can be used to extract the expected frequency dependence.  This can be compared to the frequency-scaling exponents of the ion heating rate, also measured in Nb traps at 295~K and at 4~K (Fig.~\ref{fig:freqdist} and~\cite{sedlacek2018}), namely 2.4(2) and 2.3(2), respectively.  Following~\cite{noel2018_arxiv}, we calculate heating-rate exponents, as predicted by the TAF model, of 1.95 at 295~K and 2.03 at 4~K.  The measured exponents differ significantly from those predicted by the model for our data, taken at room temperature and below, but more precise measurements of the frequency scaling over the entire temperature range of interest would be required to constrain the model further.  We note however that the temperature dependence is not predicted independently for the TAF model, and so it is difficult to completely validate it with ion heating-rate data alone.  One would ideally require a separate measure of the fluctuator energy-scale distribution from which the temperature dependence can be predicted~\cite{PhysRevLett.43.646}.

\section{Acknowledgments}

We thank Vladimir Bolkhovsky for niobium trap fabrication, George Fitch for layout assistance, and Peter Murphy, Chris Thoummaraj, and Karen Magoon for assistance with chip packaging.  We also thank Libby Shaw for help with interpretation of XPS data.  We additionally thank K.~McKay and J.~Wu for helpful comments on the manuscript.  This work made use of the MRSEC Shared Experimental Facilities at MIT, supported by the National Science Foundation under award number DMR-14-19807.  Electroplated Au traps were fabricated in the Boulder Microfabrication Facility at NIST.  This work was sponsored by the Assistant Secretary of Defense for Research and Engineering under Air Force contract number FA8721-05-C-0002. Opinions, interpretations, conclusions, and recommendations are those of the authors and are not necessarily endorsed by the United States Government.  This paper is a partial contribution of NIST and is not subject to US copyright.

\bibliography{ESIM_Bib_resub}

\end{document}